\documentclass[conference]{IEEEtran}
\IEEEoverridecommandlockouts
\usepackage{cite}
\usepackage{amsmath,amssymb,amsfonts}
\usepackage{algorithmic}
\usepackage{multirow}
\usepackage{graphicx}
\usepackage{cite}
\usepackage{hyperref}
\usepackage{cleveref}
\usepackage{textcomp}
\usepackage{xcolor}
\def\BibTeX{{\rm B\kern-.05em{\sc i\kern-.025em b}\kern-.08em
    T\kern-.1667em\lower.7ex\hbox{E}\kern-.125emX}}
\begin{document}

\title{
    Event-based Neural Spike Detection Using Spiking Neural Networks for Neuromorphic iBMI Systems
    \thanks{
        This research was supported by Korean Institute for Advancement of Technology(KIAT) grant funded by the Korea Government(MOTIE) (RS-2024-00435693, Human Resource Development Program for Industrial Innovation(Global)), the National Research Foundation of Korea (NRF) grant funded by the Korea government(MSIT) (No. RS-2024-00345732), the BK21 FOUR Project, and the Research Grants Council of the Hong Kong Special Administrative Region, China (Project No. CityU 11200922).
        
        ${\dagger}$ The first and second authors contributed equally.
    }
}

\author{
    \IEEEauthorblockN{
        Chanwook Hwang$^{1,\dagger}$, 
        Biyan Zhou$^{2,\dagger}$, 
        Ye Ke$^{2}$, 
        Vivek Mohan$^{3,4}$,
        Jong Hwan Ko$^{1}$,
        Arindam Basu$^{2,\ast}$
    }
    \IEEEauthorblockA{
        \textsuperscript{1}Department of Electrical and Computer Engineering, Sungkyunkwan University, South Korea\\
        \textsuperscript{2}Department of Electrical Engineering, City University of Hong Kong, Hong Kong\\
        \textsuperscript{3}Imperial College London \textsuperscript{4}Imperial Global Singapore\\
        Email: ghkdcks12@g.skku.edu, 
        byzhou4-c@my.cityu.edu.hk, 
        arinbasu@cityu.edu.hk
    }
}
\maketitle
\begin{abstract}
Implantable brain-machine interfaces (iBMIs) are evolving to record from thousands of neurons wirelessly but face challenges in data bandwidth, power consumption, and implant size. We propose a novel Spiking Neural Network Spike Detector (SNN-SPD) that processes event-based neural data generated via delta modulation and pulse count modulation, converting signals into sparse events. By leveraging the temporal dynamics and inherent sparsity of spiking neural networks, our method improves spike detection performance while maintaining low computational overhead suitable for implantable devices. Our experimental results demonstrate that the proposed SNN-SPD achieves an accuracy of 95.72\% at high noise levels (standard deviation 0.2), which is about 2\% higher than the existing Artificial Neural Network Spike Detector (ANN-SPD). Moreover, SNN-SPD requires only 0.41\% of the computation and about 26.62\% of the weight parameters compared to ANN-SPD, with zero multiplications. This approach balances efficiency and performance, enabling effective data compression and power savings for next-generation iBMIs.
\end{abstract}


\section{Introduction}
Implantable brain-machine interfaces (iBMIs) have seen significant advances in recent years, transitioning from tools for studying brain electrophysiology to devices capable of restoring lost sensory and motor functions in individuals with disabilities \cite{iBMI_example_1}. Traditional iBMI systems often involve wired connections and low electrode counts, which limit their practicality and scalability for real-world applications. Next-generation iBMIs are envisioned to be wireless and capable of recording from thousands of neurons simultaneously, thereby enhancing their performance and functionality \cite{iBMI_Neuralink}.

However, dealing with Moore's law-like scaling of the number of electrodes that record neurons creates challenges related to data transmission bandwidth, power consumption, and implant size. Transmitting raw neural data at high sampling rates can result in data rates of several gigabytes per second, which is impractical for wireless implants due to power and bandwidth constraints \cite{iBMI_challenge}. Efficient data compression techniques are therefore essential to reduce the amount of data transmitted without compromising critical neural information required for downstream processing tasks such as decoding and control.

A promising approach to address these challenges is spike detection. Since neural signals are believed to carry information in spikes, focusing on detecting and transmitting solely these spikes enables 100-1000$\times$ data rate compression \cite{SPD_Overview}. This approach not only reduces power consumption by decreasing the data rate required to transmit data but can also help eliminate background noise. Spike detection is implemented as a circuit known as a spike detector (SPD) on the implant device to distinguish spikes, ensuring that only relevant neural information is communicated.

Typical SPD implementations detect spikes when the amplitude or energy of the signal is larger than a certain threshold \cite{SPD_Thr, SPD_NEO}. Based on these methods, a variety of other SPDs have been proposed, including hardware-efficient adaptive noise determination \cite{SPD_AdaNoise}, firing rate-based threshold adaptation \cite{SPD_FireRate_1, SPD_FireRate_2, SPD_FireRate_3} and machine learning-based spike detection \cite{SPD_ML_1, SPD_ML_2}. However, despite the potential for significant data rate reduction, these methods still process conventional, Nyquist-rate real-valued neural signals, which require substantial computational resources within the SPD. Additionally, systems that transmit the original signal upon spike detection continue to demand high bandwidth \cite{SPD_Overview}. Consequently, there is little improvement in terms of implant size, power consumption, and memory usage—factors that are critical for the viability of next-generation iBMIs \cite{Vivek_2023}.

\begin{figure*}[t]
    \centering
    \includegraphics[width=0.9\textwidth]{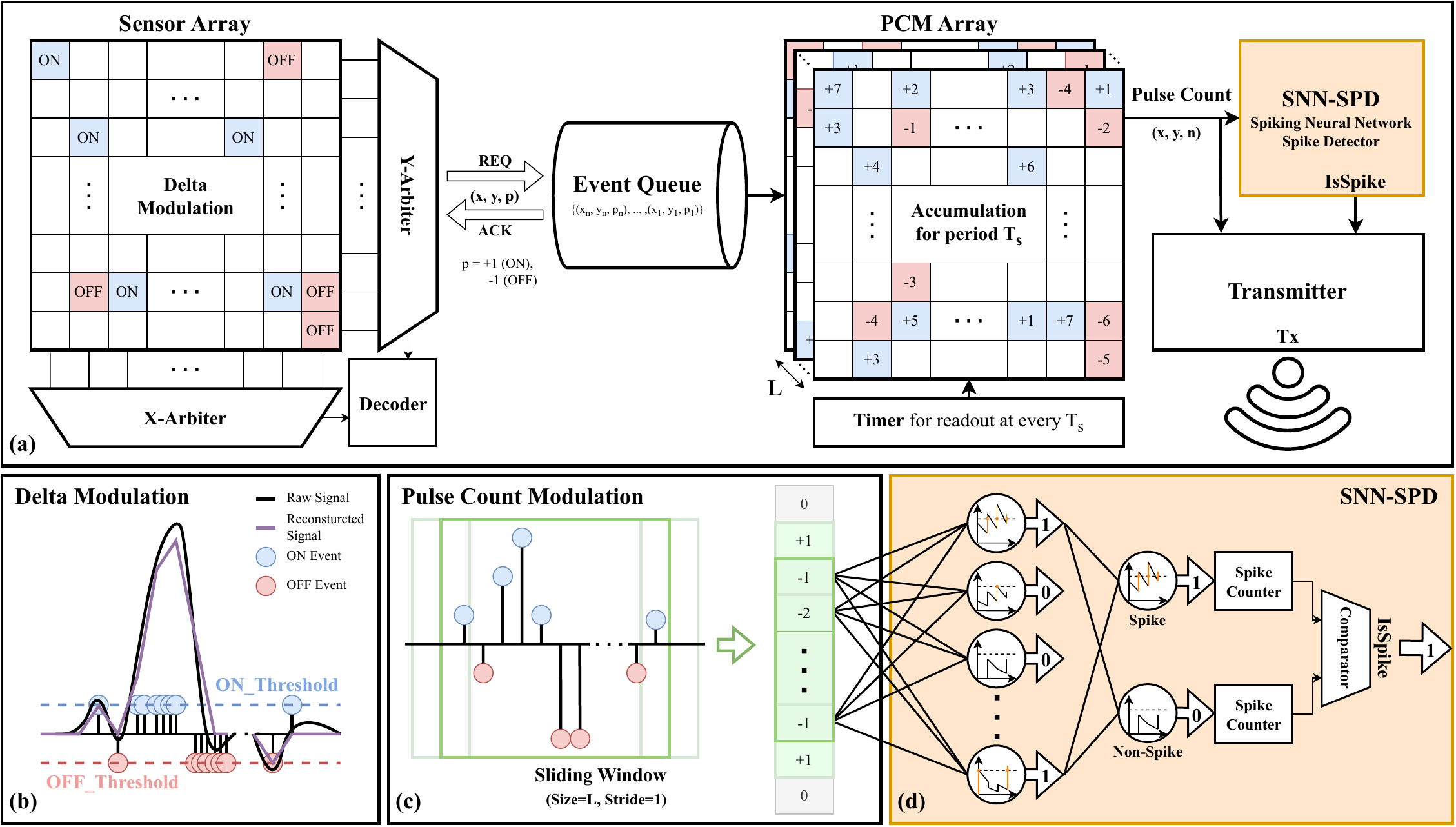}
    \vspace{-5 pt}
    \caption{(a) Spiking neural network spike detector (SNN-SPD) detects spikes from a pulse count signal generated by the iBMI sensors. (b) Delta Modulation employed for the event data generation, data is modulated to $\pm$1 when the amplitude change is greater than a threshold. (c) Pulse count modulation (PCM) is used to accumulate delta modulation values for each period $T_s$. (d) SNN-SPD detects the spikes from the pulse count signal array using a sliding window.}
    \label{fig:overall_system}
\vspace{-10 pt}
\end{figure*}

In contrast, neuromorphic approaches utilizing event-based sensors offer a promising alternative by representing neural signals as streams of asynchronous integer events. Unlike conventional frame-based sensors that generate data at fixed intervals, event-based sensors produce data only when significant changes occur in the input signal \cite{EventSensor}. In the context of neural recording, event-based neural sensors (EBNS) generate events when the amplitude of the neural signal changes significantly, which fundamentally offers data compression, generally achieving a compression ratio of 15-20$\times$ \cite{Vivek_2023}. Despite the advantages of event-based neuromorphic sensors in reducing data rates and resource consumption, spike detection in these systems presents new challenges. Conventional spike detection methods for EBNS typically require reconstructing the continuous neural signal from the event stream before applying the spike detection algorithm \cite{Vivek_2023, Event_SPD_1, Event_SPD_2}. This reconstruction step negates the benefits of data compression and introduces additional computational overhead.

Recent studies have explored alternative SPDs directly on event-based data to address these issues. For instance, the event count-based spike detector (Ev-SPD) and the artificial neural network spike detector (ANN-SPD) are proposed \cite{mohan2024hybrid}. Ev-SPD detects spikes based on the number of times an event occurs, and although its detection performance is slightly inferior, it is very computationally efficient, so it was proposed for on-implant implementation. On the other hand, ANN-SPD is a simple neural network consisting of two fully connected layers that receives event data as input to detect spikes, which has better detection performance than Ev-SPD, but was proposed for off-implant wearable devices due to the amount of computation of the neural network.

In this work, we propose a novel SNN-SPD that operates directly on event-based neural data using a spiking neural network (SNN) to enhance on-implant spike detection performance. By leveraging the temporal dynamics and inherent sparsity of SNNs, our method improves spike detection accuracy with 95.72\% for a noise level of 0.2, while maintaining low computational overhead (only 0.41\% of ANN-SPD) suitable for implantable devices. Unlike previous approaches such as Ev-SPD, which is computationally efficient but less accurate, or ANN-SPD, which is accurate but impractical for on-implant implementation due to its computational demands, our SNN-SPD strikes a balance between efficiency and performance. This advancement eliminates the need for signal reconstruction and enables effective data compression (361.96$\times$ than raw signal), paving the way for next-generation neuromorphic iBMIs capable of real-time neural processing.

\section{Methodology}

\subsection{Event Data Generation}
To efficiently represent neural signals for implantable devices, pulse count modulation (PCM) is implemented to convert continuous neural signals into sparse, event-based data. Unlike the existing method of accumulating the ON/OFF pulse counts on different channels, ON counts are represented as positive and OFF counts are represented as negative on one channel to further reduce the input size of the SPD \cite{mohan2024hybrid, Vivek_2023}.

\subsubsection{Delta Modulation}
Delta modulation is a method of converting a continuous analog signal into a discrete pulse train. As shown in Fig.\ref{fig:overall_system}(b), if the change in the signal amplitude change is greater than the ON/OFF\_Threshold, it is converted to a $\pm1$ pulse. Each pixel in the sensor array of Fig.\ref{fig:overall_system}(a) contains such a delta modulation circuit. The pulse can be mathematically defined as Eq.\eqref{eqn:pulse_generation} and after each pulse, the reference voltage is updated as Eq.\eqref{eqn:V_ref_update}:
\begin{equation} 
    \label{eqn:pulse_generation}
    p(t) = \begin{cases} 
        +1, & \text{if } V_{\text{in}}(t) - V_{\text{ref}}(t) \geq Th_{ON}, \\
        -1, & \text{if } V_{\text{in}}(t) - V_{\text{ref}}(t) \leq Th_{OFF}
    \end{cases} 
\end{equation}
\begin{equation} 
    \label{eqn:V_ref_update}
    V_{\text{ref}}(t+\delta) = V_{\text{ref}}(t) + Th_{ON/OFF} \cdot p(t) 
\end{equation}
Practical implementations do not modify $V_{ref}$ but rather reset the signal to a baseline on every level crossing\cite{corradi_adm}.

\subsubsection{Pulse Count Modulation (PCM)}
The ON/OFF pulses are accumulated in pulse count bins (n) of duration $T_s$. The PCM value $n_{x,y}$ for channel $(x,y)$ is calculated as in Eq.\eqref{eqn:PCM}:
\begin{equation}
    \label{eqn:PCM}
    n_{x,y}(t=kT_s) = \int_{t=(k-1)T_s}^{kT_s}p_{x,y}(t)dt
\end{equation}
An example of this calculation is shown in Fig.\ref{fig:overall_system}(a). Here onward, we refer to $n_{x,y}(kT_s)$ as $n_{x,y}(k)$ for simplicity, emphasizing the discrete time nature of the signal.
\subsection{Spiking Neural Network Spike Detector (SNN-SPD)}

The SNN model consists of $C$ fully connected layers with leaky integrate-and-fire (LIF) neurons. The number of neurons in layer $j$ is denoted by $n_j$ for $1\leq j\leq C$. The membrane potential ($V_{mem}$) leakage of the LIF neuron is set as ($\beta=0.5$) while the threshold is set as ($V_{thr}=1$). The equation for the k-th LIF neuron in the (j+1)-th layer is shown in Eq.\eqref{eqn:LIF} \cite{snnTorch}:
\begin{align}
    V_{mem,k}^{j+1}(t+1) &= {\beta}V_{mem,k}^{j+1}(t) + I_{in,k}^{j+1}(t+1)\notag\\
    S_k^{j+1}(t+1) &= \begin{cases}
    1, & \text{ if } V_{mem,k}^{j+1}(t+1) > V_{thr} \\
    0, & \text{ otherwise } \notag\\
    \end{cases}\\
    I_{in,k}^{j+1}(t+1) &= \begin{cases}
    \sum\limits_{i=0}^{k}w_{ik}^{j}n_{i}^{j}(t+1), & \text{ if j = 0 } \\
    \sum\limits_{i=0}^{n_{j}}w_{ik}^{j}S_{k}^{j}(t+1), & \text{ otherwise }\\
    \end{cases}
\label{eqn:LIF}
\end{align}
Note that there is no reset mechanism of the $V_{mem}$ when a spike occurs, $V_{mem}$ decays naturally due to the resistive leakage. This mode is used since it was found to perform better  than LIF with reset due to the better temporal feature extraction ability of the SNN. Resetting of $V_{mem}$ to 0 after a spike occurs reduced accuracy by $0.34\%$. Further, $C=2$ was found to be an optimal depth of the network to balance accuracy and computational cost. Adding another hidden layer of size 4 increased accuracy by only $0.01\%$ but raised the false detection rate by $0.0004$ and added 41 more weight parameters. The number of neurons in the last layer is set to $n_C=2$, since this was found to be more robust than choosing $n_C=1$ and comparing $V_{mem}$ with a reference to denote detection. 

\subsection{SNN modes}
To demonstrate the temporal feature extraction capabilities of the SNN-SPD, we train the model in two ways: \textbf{Non-Stream} and \textbf{Stream}. In the Non-Stream mode, $V_{mem}$ is reset to zero after each sliding window. In the Stream mode, $V_{mem}$ is maintained across windows without resetting.

\subsubsection{\textbf{Non-Stream}}
In this mode, the SNN processes each input independently without temporal accumulation.
\begin{itemize}
    \item Training and Testing: The PCM train is divided into segments of length $L$, matching the SNN's input size. Each segment is fed into the SNN, and $V_{mem}$ is reset before processing each one. The SNN generates a spike for each output neuron; the neuron that fires determines the prediction for that input.
\end{itemize}

\subsubsection{\textbf{Stream}}
In the Stream mode, $V_{mem}$ is maintained across inputs to leverage temporal features.
\begin{itemize}
    \item Training: We cannot use the entire continuous data for training due to the highly imbalanced classification problem (spikes are sparse compared to non-spikes) and the need for labeling each data point. To address this, we slice the PCM train into larger segments of length $2L-1$. A sliding window of length $L$ with stride 1 moves across each segment, feeding data into the SNN without resetting $V_{mem}$. The SNN generates spike trains for each output neuron. The neuron with more spikes over the segment determines the prediction.
    \item Testing: The full PCM train is fed into the SNN using the same sliding window approach, maintaining $V_{mem}$ throughout. The prediction for each window is determined by the neurons with the highest spike count by counting the number of spikes in $L$ previous windows for each output neuron.
\end{itemize}

\section{Results}

\begin{table}[t]
\caption{Characteristic Comparison of SPDs}
\vspace{-5 pt}
\label{tbl:character_comparison}
\centering
\resizebox{\columnwidth}{!}{%
\begin{tabular}{|l|ccccc|}
\hline\hline
\multicolumn{1}{|c|}{\textbf{Factors}} & \multicolumn{1}{c|}{\textbf{APM} \cite{Vivek_2023}} & \multicolumn{1}{c|}{\textbf{PCM} \cite{Vivek_2023}} & \multicolumn{1}{c|}{\textbf{Ev-SPD} \cite{mohan2024hybrid}} & \multicolumn{1}{c|}{\textbf{ANN-SPD} \cite{mohan2024hybrid}} & \textbf{SNN-SPD} \\ \hline\hline
\textbf{Input Data} & \multicolumn{2}{c|}{\begin{tabular}[c]{@{}c@{}}Reconstructed signal\\from Events\end{tabular}} & \multicolumn{1}{c|}{\begin{tabular}[c]{@{}c@{}}2-Channel\\PCM Events\end{tabular}} & \multicolumn{1}{c|}{\begin{tabular}[c]{@{}c@{}}2-Channel\\PCM Frame\end{tabular}} & \begin{tabular}[c]{@{}c@{}}1-Channel\\PCM Events\end{tabular} \\ \hline
\textbf{\begin{tabular}[c]{@{}l@{}}Calibration\\ or Training\end{tabular}} & \multicolumn{2}{c|}{Yes} & \multicolumn{1}{c|}{No} & \multicolumn{1}{c|}{Yes} & Yes \\ \hline
\textbf{Complexity} & \multicolumn{2}{c|}{High} & \multicolumn{1}{c|}{Low} & \multicolumn{1}{c|}{High} & Medium \\ \hline
\textbf{\begin{tabular}[c]{@{}l@{}}Compression\\ Ratio\end{tabular}} & \multicolumn{1}{c|}{12.97} & \multicolumn{1}{c|}{12.66} & \multicolumn{3}{c|}{361.96} \\ \hline
\textbf{\begin{tabular}[c]{@{}l@{}}Operation\\ Location\end{tabular}} & \multicolumn{2}{c|}{Offline} & \multicolumn{1}{c|}{On-implant} & \multicolumn{1}{c|}{\begin{tabular}[c]{@{}c@{}}Off-implant\\ (Wearable)\end{tabular}} & On-implant \\ \hline
\textbf{\begin{tabular}[c]{@{}l@{}}Computation\\ Count\end{tabular}} & \multicolumn{3}{c|}{\multirow{2}{*}{-}} & \multicolumn{1}{c|}{100\%} & 0.41\% \\ \cline{1-1} \cline{5-6} 
\textbf{\begin{tabular}[c]{@{}l@{}}Model Size\end{tabular}} & \multicolumn{3}{c|}{} & \multicolumn{1}{c|}{100\%} & 26.62\% \\ \hline
\textbf{Accuracy} & \multicolumn{1}{c|}{0.92-0.94} & \multicolumn{1}{c|}{0.89-0.99} & \multicolumn{1}{c|}{0.92-0.99} & \multicolumn{1}{c|}{0.95-0.99} & 0.96-0.99 \\ \hline\hline
\end{tabular}%
}
\vspace{-10 pt}
\end{table}

\subsection{Dataset}
The dataset was constructed using three spike shapes from neocortex and basal ganglia recordings \cite{dataset}. To simulate background neuronal noise, Gaussian noise with standard deviations of 0.05, 0.1, 0.15, and 0.2 was added. A total of 60 seconds of data was sampled at 24 kHz, with spikes occurring according to a Poisson distribution at a firing rate of 20 Hz. The recordings were converted to PCM events by accumulating ON/OFF pulses ($\pm$1) between each time step, resulting in an event sparsity of 0.2 as in \cite{Vivek_2023, mohan2024hybrid}. These PCM events were used directly as input for the SNN-SPD.

\subsection{Evaluation Metrics}
SPDs are evaluated in two directions: detection performance and operational efficiency.

\subsubsection{Detection Performance}
To evaluate the performance of the SPDs, the detected spike times are compared with the ground truth spike times provided in the dataset. A detected spike is considered a true positive (TP) if it occurs within a window of $t_{spk} \pm \delta t$, where $t_{spk}$ is the ground truth spike time and $\delta t$ is set to half of the average spike duration (i.e., 0.5 ms). Incorrect detections that do not correspond to a ground truth spike within this window are counted as false positives (FP), and missed detections where no spike is detected within the window of a ground truth spike are counted as false negatives (FN). \textbf{Sensitivity (S)}, \textbf{false detection rate (FDR)}, and \textbf{accuracy (A)} are calculated using Eq.\eqref{eqn:performance_metrics} and used for spike detection performance evaluation.
\begin{equation}
    \label{eqn:performance_metrics}
    S=\frac{TP}{TP+FN}; FDR=\frac{FP}{TP+FP}; A=\frac{TP}{TP+FP+FN}
\end{equation}

\subsubsection{Operation Efficiency}
The following five metrics are selected to compare the operational efficiency of ANN-SPD and SNN-SPD. 
\begin{itemize}
    \item \textbf{Multiplication count}: Multiplication between input features and weight parameters is essential for ANNs with real number-based operations. On the other hand, SNNs with event-driven integer-based operations can convert multiplication and accumulation (MAC) to multiple accumulations in an asynchronous system \cite{SNN_Async}.
    \item \textbf{Accumulation count}: The rate at which the MAC is replaced by accumulations is determined by the pulse count of the input feature. For example, for an input with a 3 pulse count, 1 MAC can be replaced by 3 accumulations. Therefore, the number of accumulation operations in SNN is calculated by multiplying the number of MAC operations by the average pulse count or spike rate.
    \item \textbf{Number of weight parameter}: The number of weight parameters the neural network has been trained on. The number is used because the bit precision of ANN and SNN is considered to be the same.
    \item \textbf{Number of output feature}: For ANN, the output feature is the result of the MAC calculation; for SNN, the output feature is the $V_{mem}$.
    \item \textbf{Interconnection}: The amount of on-chip interconnection for the value that should be passed to the next layer. For ANN, the ReLU activated feature is passed, which is equal to the number of features multiplied by 32 bits, and for SNN, the spike (0 or 1) is passed, which is equal to the number of features multiplied by 1 bit.
\end{itemize}

\subsection{Detection Performance Comparison}
\begin{figure}[!t]
    \centering
    \includegraphics[width=\columnwidth]{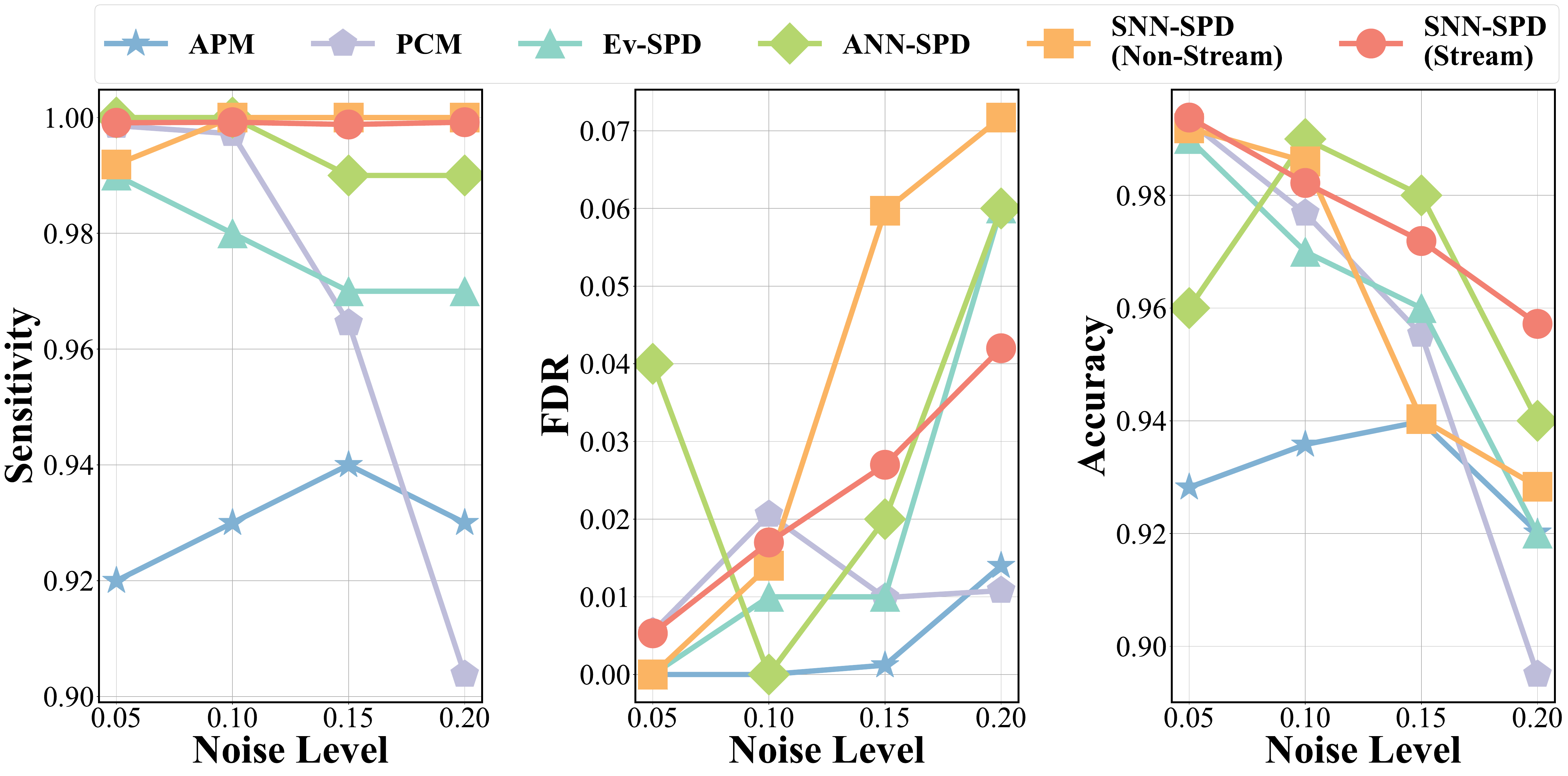}
    \vspace{-20 pt}
    \caption{Spike detection performance comparison between PCM, APM, Ev-SPD, ANN-SPD, and SNN-SPDs (Stream and Non-Stream)}
    \label{fig:acccompare}
\vspace{-10 pt}
\end{figure}

The models were trained based on the PyTorch and snnTorch frameworks \cite{pyTorch, snnTorch}. The optimizer is chosen as Adam with a learning rate of $0.0005$, and cross-entropy spike count loss from snnTorch is determined as the loss function. As a surrogate function to solve the in-differentiability of spikes in SNN, arc tangent is used \cite{neftci2019surrogate}. Fig.\ref{fig:acccompare} and \Cref{tbl:character_comparison} compare the performance of SPDs. SNN-SPD (Stream) consistently demonstrates high S and A across all noise levels due to its effective temporal feature extraction and its inherent robustness. ANN-SPD and SNN-SPD (Non-Stream) perform comparably to SNN-SPD (Stream) at low noise levels but experience a significant performance drop ($\approx$2\%) at higher noise levels. The increased FDR in SNN-SPD (Non-Stream) suggests that without temporal accumulation, it is more sensitive to spikes. Ev-SPD performs well at low noise levels with a lower FDR but sees a sharp decline in S and A as noise increases. At a noise level of 0.2, SNN-SPD (Stream) achieves the highest A of 95.72\%, about 3\% higher than SNN-SPD (Non-Stream) and 2\% higher than ANN-SPD. The compression ratio is calculated \cite{Mohan_2025_NCE} and indicates that \Cref{tbl:character_comparison}, SNN-SPD achieves the highest compression ratio of 361.96 compared to the raw data. This is 28.5$\times$ higher than APM/PCM, with only 0.41\% of the computation and 26.62\% model size compared to ANN-SPD.

\subsection{Operation Efficiency Comparison}

\begin{table}[t]
\caption{Operation efficiency comparison}
\vspace{-5 pt}
\label{tbl:efficiency_comparison}
\centering
\resizebox{0.8\columnwidth}{!}{%
\begin{tabular}{|l|c|c|}
\hline\hline
\multicolumn{1}{|c|}{\textbf{Metrics}} & \textbf{ANN-SPD} \cite{mohan2024hybrid} & \textbf{SNN-SPD} \\ \hline\hline
\textbf{Model Architecture} & 2$\times$47-32-2 & 1$\times$24-16-2 \\ \hline\hline
\textbf{Multiplication Count} & 3,072 & 0 \\ \hline
\textbf{Accumulation Count} & 3,072 & 25.23 \\ \hline
\textbf{\# Weight Parameter} & 1,570 & 418 \\ \hline
\textbf{\# Output Feature} & 34 & 18 \\ \hline
\textbf{Interconnection (bit)} & 1,088 & 18 \\ \hline\hline
\end{tabular}%
}
\vspace{-10 pt}
\end{table}

\Cref{tbl:efficiency_comparison} compares the efficiency of the models. Ev-SPD is the simplest, requiring no training and easy implementation, but its effectiveness is limited in complex environments. The focus is on comparing ANN-SPD and SNN-SPD. SNN-SPD offers significant hardware implementation advantages due to its smaller model size and elimination of multiplications in asynchronous systems. It requires only 1/120 of the accumulations and about 1/4 of the weight parameters to achieve 2\% higher accuracy than ANN-SPD. Moreover, SNN-SPD (Non-Stream) is even more efficient as it does not need to maintain $V_{mem}$ at each time step.

\section{Conclusion}
We introduced an SNN-SPD that enhances on-implant spike detection by processing event-based data with spiking neural networks, improving accuracy while significantly reducing computational demands. Our method outperforms existing SPDs in both detection performance and efficiency. Specifically, SNN-SPD achieves 95.72\% accuracy at a noise level of 0.2, approximately 2\% higher than ANN-SPD, while requiring only 25.23 accumulations compared to 3,072 MAC operations in ANN-SPD, reducing computational load by over 99\%. This advancement enables effective data compression and power savings, making it suitable for on-implant implementation in next-generation neuromorphic iBMIs. Future work will focus on hardware implementation of the SNN-SPD and validation with real biological neural recordings to further establish its applicability in practical iBMI systems.

\clearpage
\clearpage

\bibliography{refs.bib}
\bibliographystyle{IEEEtran} 

\end{document}